\documentstyle[pra,aps,epsfig,twocolumn]{revtex}

\begin{document}
\draft
\title{Radiative recombination of bare Bi$^{83+}$:\\
 Experiment versus theory}
\author{A.~Hoffknecht, C.~Brandau, T.~Bartsch, C.~B\"{o}hme, H.~Knopp, S.~Schippers and A.~M\"uller}
\address{Institut f\"ur Kernphysik, Universit\"at Giessen, D-35392 Giessen, Germany}
\author{C.~Kozhuharov, K.~Beckert, F.~Bosch, B.~Franzke, A.~Kr\"{a}mer, P.H.~Mokler, F.~Nolden, M.~Steck and Th.~St\"{o}hlker}
\address{Gesellschaft f\"{u}r Schwerionenforschung (GSI), D-64291 Darmstadt, Germany}
\author{Z.~Stachura}
\address{Institute for Nuclear Physics, 31-342 Krak\'ow, Poland}

\date{\today}
\maketitle

\begin{abstract}
Electron-ion recombination of completely stripped Bi$^{83+}$ was
investigated at the Experimental Storage Ring (ESR) of the GSI in
Darmstadt. It was the first experiment of this kind with a bare
ion heavier than argon. Absolute recombination rate coefficients
have been measured for relative energies between ions and
electrons from 0 up to about 125~eV. In the energy range from
15~meV to 125~eV a very good agreement is found between the
experimental result and theory for radiative recombination (RR).
However, below 15~meV the experimental rate increasingly exceeds
the RR calculation and at E$_{rel}$= 0~eV it is a factor of 5.2
above the expected value. For further investigation of this
enhancement phenomenon the electron density in the interaction
region was set to 1.6$\times$10$^{6}$~cm$^{-3}$, 3.2$\times
10^{6}$~cm$^{-3}$ and 4.7$\times$10$^{6}$~cm$^{-3}$. This
variation had no significant influence on the recombination rate.
An additional variation of the magnetic guiding field of the
electrons from 70~mT to 150~mT in steps of 1~mT resulted in
periodic oscillations of the rate which are accompanied by
considerable changes of the transverse electron temperature.
\end{abstract}

\section{Introduction}\label{sec:intro}
Recombination  between electrons and highly charged ions plays an
important role in different areas of modern physics. The basic
two- and three-body  recombination processes are of very
fundamental nature and thus provide an excellent testing ground
for collision theory and atomic structure calculations. Cross
sections and rate coefficients of these processes are needed for
the understanding of astrophysical and fusion plasmas and also
provide useful information for applications in accelerator physics
\cite{rec1}. In particular, beam losses in ion storage rings by
electron-ion recombination during electron cooling can post harsh
limits to the handling and the availability of ions for further
experiments.

During the last decade electron-ion recombination has been
extensively investigated. Merged-beams experiments using storage
ring coolers and single-pass electron targets at accelerators have
provided excellent conditions for a new generation of
recombination measurements. In these experiments an incident ion
beam is merged with a cold beam of  electrons over a distance of
typically 50 to 250~cm depending on the specific electron beam
device. By choosing the appropriate accelerator facility, ions of
most elements in all possible charge states can be investigated
nowadays.

Free electrons can be captured by  ions via several different
mechanisms. The main recombination channel for a bare ion is
radiative recombination (RR)
\begin{equation}
e + A^{q+} \to A^{(q-1)+} + h\nu.
\end{equation}
RR is the direct capture of a free electron by an ion A$^{q+}$
where the excess energy and momentum are carried away by a photon.
After the capture, which is inverse to photoionization, the
electron can be in an excited state and there will be further
radiative transitions within the ion until the electron has
reached the lowest accessible energy level. RR is a non-resonant
process with a diverging cross section at zero center-of-mass
(c.m.) energy which continuously decreases towards higher
c.m.~energies.

Another recombination mechanism  possible for a bare ion is
three-body recombination (TBR)
\begin{equation}
\label{equ:tbr}
 A^{q+} + e + e \to  A^{(q-1)+} + e'
\end{equation}
where the  excess energy and momentum are carried away by a second
electron. This process is important at high electron densities and
very low center-of-mass energies between electrons and ions.

The pioneering experiment on radiative recombination of bare ions
was performed by Andersen et al.~\cite{andersen1} in 1990.
Absolute rate coefficients were measured for C$^{6+}$ with a
merged-beam technique finding a reasonably good agreement between
experiment and theory in the investigated energy range from
E$_{rel}$= 0  to 1~eV where E$_{rel}$ is the relative energy
between the ensembles of electrons and ions in the interaction
region (for the definition of E$_{rel}$ see section
\ref{sec:evaluation}).

In a number of consecutive measurements different bare ions
(D$^+$,He$^{2+}$,C$^{6+}$,N$^{7+}$,Ne$^{10+}$,Si$^{14+}$,Ar$^{18+}$)
\cite{gao1,wolf1,uwira1,gao2} have been investigated at several
facilities. The measured rate coefficients have been in accordance
with a theoretical approach to RR by Bethe and Salpeter
\cite{bethe} for relative energies $E_{rel} \geq$ 0.01~eV. The
experiments were limited by counting statistics to an energy range
of only a few eV. Recently an experiment with Cl$^{17+}$-ions
revealed excellent agreement between the measured rate and RR
theory for relative energies from 0.01~eV up to 47~eV
\cite{hoffknecht2}. However, in all of these measurements strong
deviations of the experimental findings from the theoretical
predictions were found at very low electron-ion relative energies
($E_{rel} \leq 0.01$~meV) . The measured rate coefficient
typically shows an additional increase towards lower energies
resulting in a so-called rate enhancement factor $\epsilon=
\alpha_{exp}/\alpha_{RR}$ at $E_{rel}$ = 0~eV of 1.6 (He$^{2+}$)
to 10 (Ar$^{18+}$) for bare ions (no enhancement observed with
D$^+$) and up to 365 for a multicharged multi-electron system like
Au$^{25+}$ \cite{hoffknecht1}. It should be noted that in contrast
to the RR cross section which diverges at $E_{rel}$ = 0~eV the
measured rate coefficient obtains a finite value due to the
experimental electron and ion velocity spreads (see section 3).

After the first observation of the so called enhancement
phenomenon in an experiment with U$^{28+}$-ions at the GSI in
Darmstadt in 1989 \cite{mueller1} and a later experimental
confirmation of that same result \cite{mitnik} this effect has
been observed over and over again in many different experiments at
different facilities. Whereas the very high rate enhancement
factors of multicharged complex ions like Au$^{25+}$, Au$^{50+}$,
Pb$^{53+}$ and U$^{28+}$ could be partly traced back to the
presence of additional recombination channels such as dielectronic
recombination (DR) \cite{hoffknecht1,uwira2,baird,gribakin} or
polarization recombination (PIR) \cite{bureyeva} the origin of the
observed discrepancies between experiment and theory for bare ions
is still unknown. Theoretical calculations applying $Molecular
Dynamics$ (MD) computer simulations \cite{zwick1,zwick2} have been
performed by {\em Spreiter et al.} \cite{spreiter}. Taking into
account the experimental conditions at the heavy ion storage ring
TSR in Heidelberg they found an increase of the local electron
density around the ion. This could possibly lead to a higher
number of recombination processes but to date the recombination of
the electrons with the ion is not included in their theoretical
description.

 For the clarification of this
phenomenon dependences of the enhancement on external experimental
parameters have been studied at different storage rings.
Variations of the electron density \cite{gao2,hoffknecht2} within
a total range from about 10$^6$ cm$^{-3}$ up to almost 10$^{10}$
cm$^{-3}$ showed little effect on the enhancement. The dependence
of the total recombination rate followed the $T_{\perp}^{-1/2}$
dependence on the mean transverse electron energy spread
$T_{\perp}$ \cite{schramm} as expected for RR alone. A systematic
study of the ion charge-state dependence of the excess rate
coefficient for a number of bare ions \cite{gao1} yielded roughly
a Z$^{2.8}$ scaling for atomic numbers 1 $\leq$ Z $\leq$ 14. The
first external parameter observed to influence the enhancement has
been the magnetic field strength in the interaction region of
electrons and ions \cite{hoffknecht1}. The experimental results
show a clear increase of the recombination rate maximum of
Au$^{25+}$ ions at $E_{rel} = 0$ with increasing magnetic field
strength. Recently this dependence has also been observed in
experiments with lithium-like F$^{6+}$ and bare C$^{6+}$-ions at
the TSR heavy ion storage ring in Heidelberg
\cite{hoffknecht3,gwinner}.

In the present measurement, which was carried out at the
Experimental Storage Ring of the GSI in Darmstadt, we extended the
range of investigated bare ions to the high-Z ion Bi$^{83+}$. It
was the first experiment of this kind with such a heavy bare ion
providing information on RR in addition to studies of the
radiative electron capture in ion-atom collisions using x-ray
spectroscopy \cite{beyer,stoehlker1,stoehlker1a}. Beside a
comparison of measured absolute rate coefficients with calculated
RR rates for relative energies from $E_{rel}$ = 0~eV to 125~eV
dependences of the recombination rate on the electron density and
the magnetic guiding field of the electron beam in the cooler have
been investigated.

The present paper is organized  as follows. After a short
description of  the theoretical approaches, the  experimental
setup and the evaluation of recombination rates are described. The
experimental results are presented  and compared with RR
calculations.

\section{Theory}
\label{sec:theory} In order to describe RR theoretically Kramers
developed a semi-classical theory already in 1923 \cite{kramers}.
A full quantum mechanical treatment was performed by Stobbe seven
years later \cite{stobbe}. In 1957, Bethe and Sal\-peter
\cite{bethe} derived an approximate formula for the RR cross
section that is identical to Kramers´ result
\begin{equation}
\label{equ:sigma} \sigma_{RR}(n,E_{cm})=(2.1 \times 10^{-22} cm^2)
\frac{E_0^2}{nE_{cm}(E_0+n^2E_{cm}).}
\end{equation}
The capture of an electron by a bare ion produces a hydrogenic
state with principal quantum number $n$. In this case E$_0$ is the
binding energy of the ground state electron in the hydrogenic ion
or atom and E$_{cm}$ is the kinetic energy in the electron-ion
center-of-mass frame. The total cross section for this process is
obtained by summing up the contributions of all accessible Rydberg
states:
\begin{equation}
\sigma_{RR}(E_{cm})= \sum_{n=1}^{n_{max}} \sigma_{RR}(n,E_{cm})
\end{equation}
where $n_{max}$ is the maximum principal quantum number that can
contribute. This number is generally limited by experimental
conditions. The approach of Bethe and Sal\-peter clearly shows the
typical features of RR cross sections: the divergence at zero
electron energy and a monotonic decrease for increasing electron
energy. However, as a semi-classical approximation
Eq.~\ref{equ:sigma} is only valid in the limit of high quantum
numbers and low electron energies, i.e.\ for $ n~\gg~1$ and
$E_{cm}~\ll~Z^2/n^2 Ry$. Since the quantum mechanical treatment of
Stobbe involves the rather tedious evaluation of hydrogenic dipole
matrix elements one often applies correction factors
$G_n(E_{cm})$, the so called Gaunt factors, to Eq.~\ref{equ:sigma}
to account for deviations from the correct quantum result at low
$n$ and high $E_{cm}$. The use of Gaunt factors is convenient
because they are either tabulated \cite{omidvar} or given in an
easy parametrization \cite{drake}. We here apply tabulated
\cite{andersen} values $k_n=~G_n(0)$ using\\

\begin{eqnarray}
\label{equ:sigmasum}
 \sigma_{RR}(E_{cm}) &=& \sum_{n=1}^{n_{max}} (2.1 \times 10^{-22} cm^2)~k_n \times \nonumber\\
 & & \frac{(Z^2Ry)^2}{nE_{cm}(Z^2~Ry+n^2E_{cm})}.
\end{eqnarray}
where $Z$ is the nuclear charge of the ion and $Ry$ = 13.6~eV is
the ground-state energy of the hydrogen atom.
Eq.~\ref{equ:sigmasum} is exact for bare ions at zero relative
energy and does not deviate by more than about 5\% from the
quantum mechanically correct hydrogenic result at the highest
energies considered in this paper.

For high-Z ions and high relative energies the non-relativistic
dipole approximation presented above is not valid. In general an
exact relativistic calculation with the inclusion of higher
multipoles is required \cite{ichihara}. Nevertheless experiments
investigating the Radiative Electron Capture (REC) into the
K-shell of high-Z ions have shown that the experimental results
are in very good overall agreement with the non-relativistic
dipole approximation \cite{stoehlker2} as long as $E_{rel}$ is
below the electron rest energy $m_ec^2$. Because of this observed
accordance Eq.~\ref{equ:sigmasum} is used for the calculations
presented in this paper.

\section{Basic relations}
\label{sec:evaluation} In the experiment cross sections at very
low center-of-mass energies $E_{cm}$ are not accessible due to the
finite velocity spread of the colliding particles. In the present
work the lowest limit of $E_{cm}$ is at about
1$\times$10$^{-5}$~eV. The measured quantity is a rate coefficient
$\alpha$ which theoretically results from a convolution of the
cross section with the velocity distribution function $f(\vec{v})$
\\
\begin{equation}
\label{eq:faltung} \alpha(v_{rel})= \int \sigma(v) v
f(v_{rel},\vec{v}) d^3v.
\end{equation}

In our experiment the ion velocity distribution is negligibly
narrow compared to that of the electrons due to the cooling of the
ion beam to a relative momentum spread below 10$^{-4}$ and due to
the large mass difference between electrons and ions. Therefore
the distribution function $f(\vec{v})$ is dominated by the
electron velocity distribution. Considering the axial symmetry of
the merged beams experiment two velocity coordinates are
sufficient to describe the distribution: $v_{\|}$ the velocity
component in beam direction and $v_{\bot}$ the velocity component
perpendicular to the beam. The appropriate velocity (or energy)
spreads are characterized by two corresponding temperatures
$T_{\|}$ and $T_{\bot}$. Due to the acceleration of the electrons
these temperatures are quite different ($T_{\|}$~$\ll$~$T_{\bot}$)
resulting in a highly anisotropic velocity distribution
$f(\vec{v})$ which is therefore often called `flattened'. Its
mathematical form is given by
\begin{eqnarray}\label{eq:fele}
 f(v_{rel},\vec{v}) & = &\frac{m_e}{2 \pi kT_{\bot}}
\exp(-\frac{m_ev_{\bot}^2}{2kT_{\bot}})   \times \nonumber\\ & &
\sqrt{\frac{m_e}{2 \pi kT_{||}}}
\exp(-\frac{m_e(v_{||}-v_{rel})^2}{2kT_{||}})
\end{eqnarray}
with $v_{rel}$ being the average longitudinal center-of-mass
velocity
\begin{equation}
v_{rel}= \mid v_{e,||}-v_{i,||}\mid /(1- v_{i,||}v_{e,||}/c^2).
\end{equation}
where $v_{e,\|}$ and $v_{i,\|}$ are the longitudinal velocity
components of the electron and ion beam in the laboratory frame,
respectively. The relative energy of electrons and ions is
\begin{equation}\label{relative}
E_{rel}= (\gamma_{rel} - 1) m_{e} c^{2}
\end{equation}
with
\begin{equation}\label{gamma}
\gamma_{rel} = \left[1- \left(v_{rel}/c \right)^{2}
\right]^{-1/2}.
\end{equation}
Under cooling conditions $v_{rel}$ is zero and, hence, also
E$_{rel}$ is zero then. For comparisons of experimental results
with theory the cross sections resulting from
Eq.~\ref{equ:sigmasum} have to be convoluted with the experimental
velocity distribution according to Eq. \ref{eq:faltung} using the
experimental temperatures $T_{||}$ and $T_{\perp}$.

\section{Experiment}
\label{sec:experiment} 
\begin{figure}
  \begin{center}
  \epsfxsize=\columnwidth
  \epsfbox{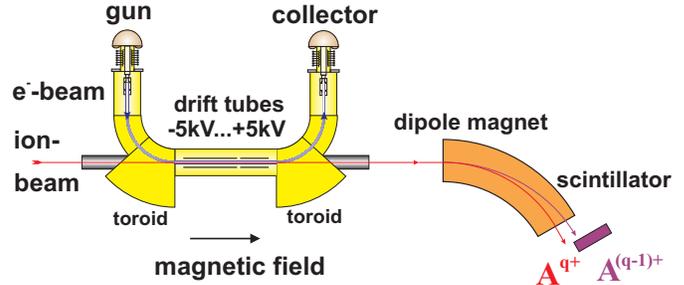}
  \end{center}
  \caption{Schematic view of the ESR electron cooler and the experimental set-up
  for recombination measurements. The cold electron beam produced in the gun is
  guided by the magnetic field and merged with the ion beam over
  a distance of 2.5~m. The electron beam is then separated from the ion beam by the
  magnetic guiding field and transferred to the collector. Recombined
  and parent ions which leave the cooler together are separated from each other
  in the first dipole magnet after the cooler. A scintillator
  detector is used to count the recombined particles.}\label{fig:cooler}
\end{figure}

The measurements have been performed at the
Experimental Storage Ring (ESR) of the Gesellschaft f\"ur
Schwerionenforschung (GSI) in Darmstadt \cite{franzke}. 295.3
MeV/u Bi$^{83+}$-ions supplied by the GSI linear accelerator
UNILAC and the heavy ion synchrotron SIS were injected into the
ESR. Only one shot of ions was sufficient to provide an ion
current of typically 400-800~$\mu$A at the beginning of a
measurement. In the storage ring the circulating Bi$^{83+}$ ions
were merged with the magnetically guided electron beam of the
electron cooler with an electron energy of 162~kV
(Fig.~\ref{fig:cooler}). Before starting a measurement, the ion
beam was cooled for several seconds until the beam profiles
reached their equilibrium widths. During the measurement the
electron energy was stepped  through a preset range of values
different from the cooling energy thus introducing non-zero mean
relative velocities $v_{rel}$ between ions and electrons. After
each voltage step a cooling interval was inserted. The whole
scheme of energy scan measurements was realized by applying
voltages from -5~kV to 5~kV to the two drift tubes surrounding the
electron and the ion beam in the interaction region. The voltages
were supplied by a system of sixteen individual power supplies
controlled by very fast high-voltage switches. This instrument has
specially been constructed for the recombination experiments
\cite{horneff}. Only 2~ms are needed by this device to switch to
and set a certain voltage with a relative precision of 10$^{-4}$.
The repetition rate of voltage settings with this precision is
limited to 40 per second.

Recombined Bi$^{82+}$ ions were counted as a function of the
electron energy on a scintillator detector located behind the
first dipole magnet downstream of the electron cooler. The dipole
magnet bends the circulating Bi$^{83+}$ ion beam onto a closed
orbit and separates the recombined Bi$^{82+}$ ions from that
orbit. In between two measurement steps of 40~ms duration, the
electron energy was always set to the cooling energy for 20~ms in
order to maintain good ion beam quality.  The experimental data
stream was continuously collected and stored after each ms. The
time-resolved measurements allowed us to observe and eliminate
drag force effects from the data in a detailed off-line analysis.
Such effects are a result of the cooling force exerted by the
electrons on the ion beam which can lead to a time-dependent shift
of the ion velocity towards the electron velocity. The friction
forces are particularly effective at relative energies close to
zero.

The kinetic energy $E_{e}$ of the electrons is defined by the
cathode voltage $U_{gun}$, the drift tube voltage $U_{drift}$ and
the space charge potential $U_{sp}$ in the interaction region.
Assuming coaxial beams it is calculated as
\begin{eqnarray}
\label{space}
 E_{e} & = &- eU_{gun} + eU_{drift} - eU_{sp}      \nonumber \\
      & =  &- eU_{gun} + eU_{drift} - \frac{I_{e}r_{e}m_{e}c^{2}}{ev_{e}}\left[1+2 \ln(b/a) \right],
\end{eqnarray}
where $r_{e}$ is the classical electron radius. The quantities $b$
= 10~cm and $a$ = 2.54~cm are the radii of the drift tube and the
electron beam, respectively. The ion beam diameter is only of the
order of a millimeter and, hence, the electron energy distribution
probed by the ions is rather flat across the ion beam. In the
present experiment we performed measurements with different
electron currents $I_{e}$ which produced space charge potentials
ranging from 17.2~V to 51.7~V.

The space charge corrected electron energy $E_e$ and the ion
energy $E_i$ are used to calculate the relative energy of
electrons and ions in the center-of-mass frame. A relativistic
transformation yields \\
\begin{eqnarray}\label{equ:rel_energy}
E_{rel} & = & \sqrt{(m_i c^2+m_e c^2)^2+} \nonumber\\ & &
\overline{2\left(E_i E_e + E_i m_e c^2 + E_e m_i c^2 - A\right)} -
B, \nonumber\\ A & = & \sqrt{E_i (E_i + 2 m_i c^2)}\sqrt{E_e(E_e +
2 m_e c^2)} \cos(\phi) \nonumber\\ B & = & (m_i c^2 + m_e c^2)
\nonumber\\
\end{eqnarray}
where $\phi$ is the angle between the electron and the ion beam
directions. According to Eq.~\ref{equ:rel_energy} the minimum
relative energy $E_{rel}$ = 0~eV cannot be reached if an angle
$\phi \neq$ 0 is present. Therefore the alignment of the beams was
optimized before the recombination experiments in order to achieve
$\phi$ = 0~mrad with an uncertainty of 0.1~mrad.

 The counting rate measured at the scanning energy $E_{meas}$
is given by
\begin{equation}\label{measrate}
    R(E_{meas}) = \frac{\alpha(E_{meas}) \eta L n_e(E_{meas}) N_i}{C \gamma^{2}} + R_{back}
\end{equation}
with $\alpha$ denoting the electron-ion recombination rate
coefficient, $\eta$ the detection efficiency of the scintillator
detector which is very close to unity, $L=2.5$~m the nominal
length of the interaction zone, $n_e$(E) the electron density at
energy $E$, $N_{\rm i}$ the number of stored ions, $C = 108.36$~m
the ring circumference and $\gamma$ the relativistic Lorentz
factor for the transformation between the c.~m.\  and the
laboratory frames. $R_{back}$ denotes the measured background rate
due to collisions with residual gas molecules. In order to extract
an absolute rate coefficient from the experimental data the
background has to be subtracted by taking into account the
counting rate
\begin{equation}\label{refrate}
   R(E_{ref}) = \frac{\alpha(E_{ref}) \eta L n_e(E_{ref}) N_i}{C \gamma^{2}} + R_{back} ,
\end{equation}
at a reference energy $E_{ref}$. Combining Eqs.~\ref{measrate} and
\ref{refrate} $\alpha$ at $E_{meas}$ has to be calculated from
\begin{eqnarray}
\label{eq:alpha} \alpha(E_{meas})& = &
\frac{(R(E_{meas})-R(E_{ref})) \gamma^2}{\eta L n_{\rm
e}(E_{meas}) N_{\rm i} / C}  \nonumber \\ & &+
\alpha(E_{ref})\frac{n_{\rm e}(E_{ref})}{n_{\rm e}(E_{meas}) }.
\end{eqnarray}
Because of RR one always has a non-zero recombination rate
coefficient $\alpha(E_{ref})$ at the reference point. Usually
$E_{ref}$ is chosen such that $\alpha(E_{ref})$ practically equals
zero, but in general one has to re-add the rate which has been
neglected by subtracting $R(E_{ref}$) from $R(E_{meas})$. In the
present experiment the reference rate has been measured at the
maximum accessible scan energy $E_{ref}$= 125~eV. According to RR
theory the rate coefficient at this energy is $\alpha(E_{ref})$=
3.6$\times$ 10$^{-10}$~cm$^{3}$~s$^{-1}$. This correction leads to
a modification of the measured rate coefficient at $E_{rel}$= 0~eV
by only 0.2 $\%$.

\begin{figure}
\begin{center}
\epsfxsize=\columnwidth \epsfbox{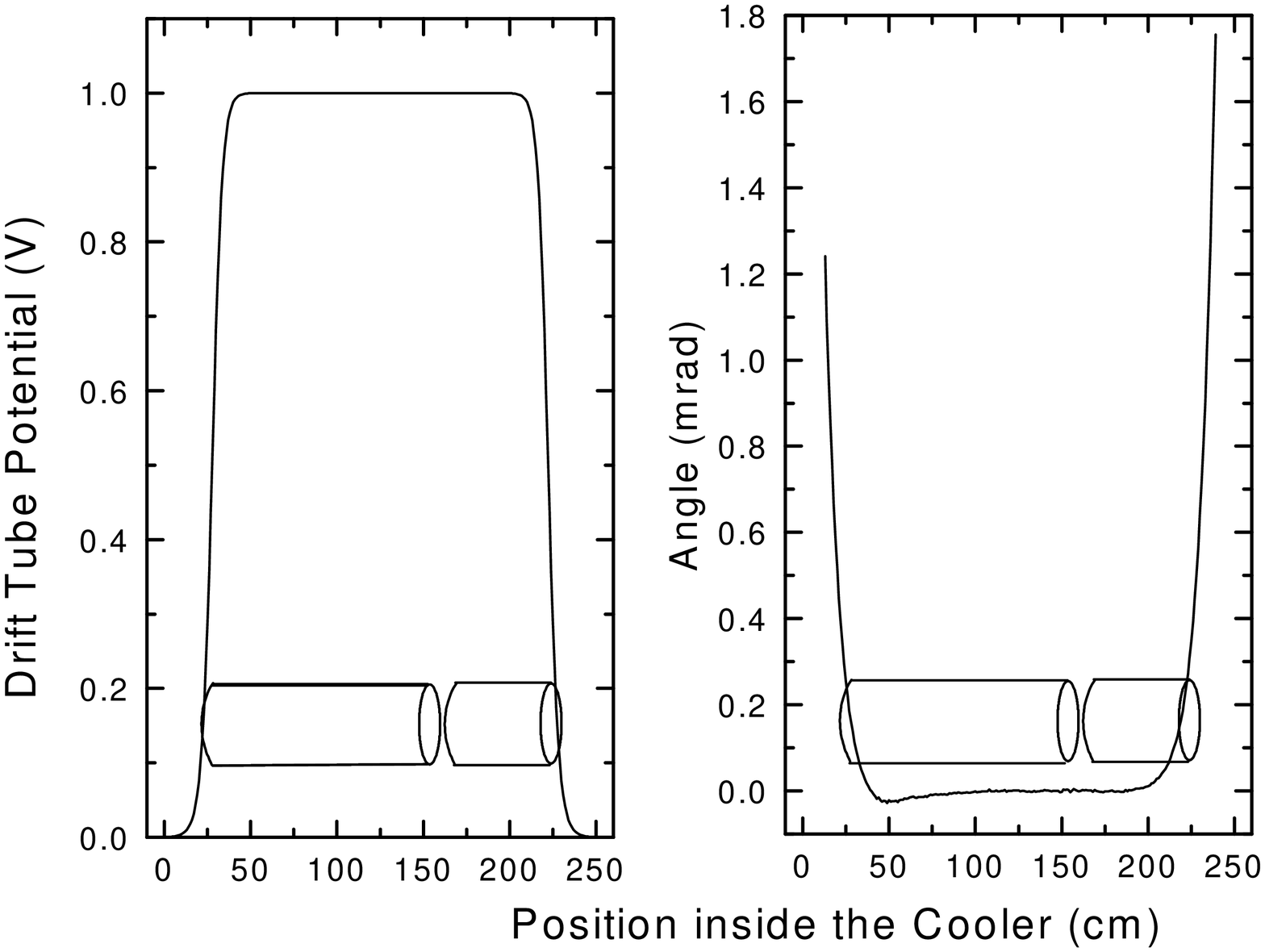}
\end{center}
  \caption{Electric potential and angle $\phi$ between
  electron and ion trajectories along the cooler axis. The dependences are displayed
 along the 2.5~m long straight section between the toroids. The
  position of the drift tubes covering 2~m of the interaction
  region
  is also indicated.
  The left panel shows the calculated electric potential
  along the cooler axis if a voltage of 1~V is applied to the drift tubes. In the right
  panel the angle $\phi$ is shown corresponding to a measurement of the
  magnetic guiding field. Here, perfect alignment of the ion beam with the geometrical
  cooler axis is assumed which is attainable by using the steerer magnets.}\label{fig:potential}
\end{figure}

Electron and ion beams are merged and demerged by bending the
electron beam in a toroidal magnetic field with a bending radius
of 120~cm. An electron beam of 2.54~cm radius is still overlapping
the ion beam for 25~cm before and after the straight overlap
section of 250~cm. The merging and demerging sections therefore
contribute to the measured counting rate. As one can see in the
left panel of Fig.~\ref{fig:potential} where the calculated
potential distribution of the drift tubes for an applied voltage
of 1~V is plotted against the position inside the cooler (along
the central axis of ion and electron beam) the influence of the
voltage applied to the drift tubes is restricted to the straight
overlap section of the cooler. Thus, the electron energy in the
toroidal sections is always the same independent of the drift tube
potential. This results in a constant contribution to the measured
counting rate which is considered by the background subtraction
procedure described above.

\begin{figure}
\begin{center}
\epsfxsize=\columnwidth \epsfbox{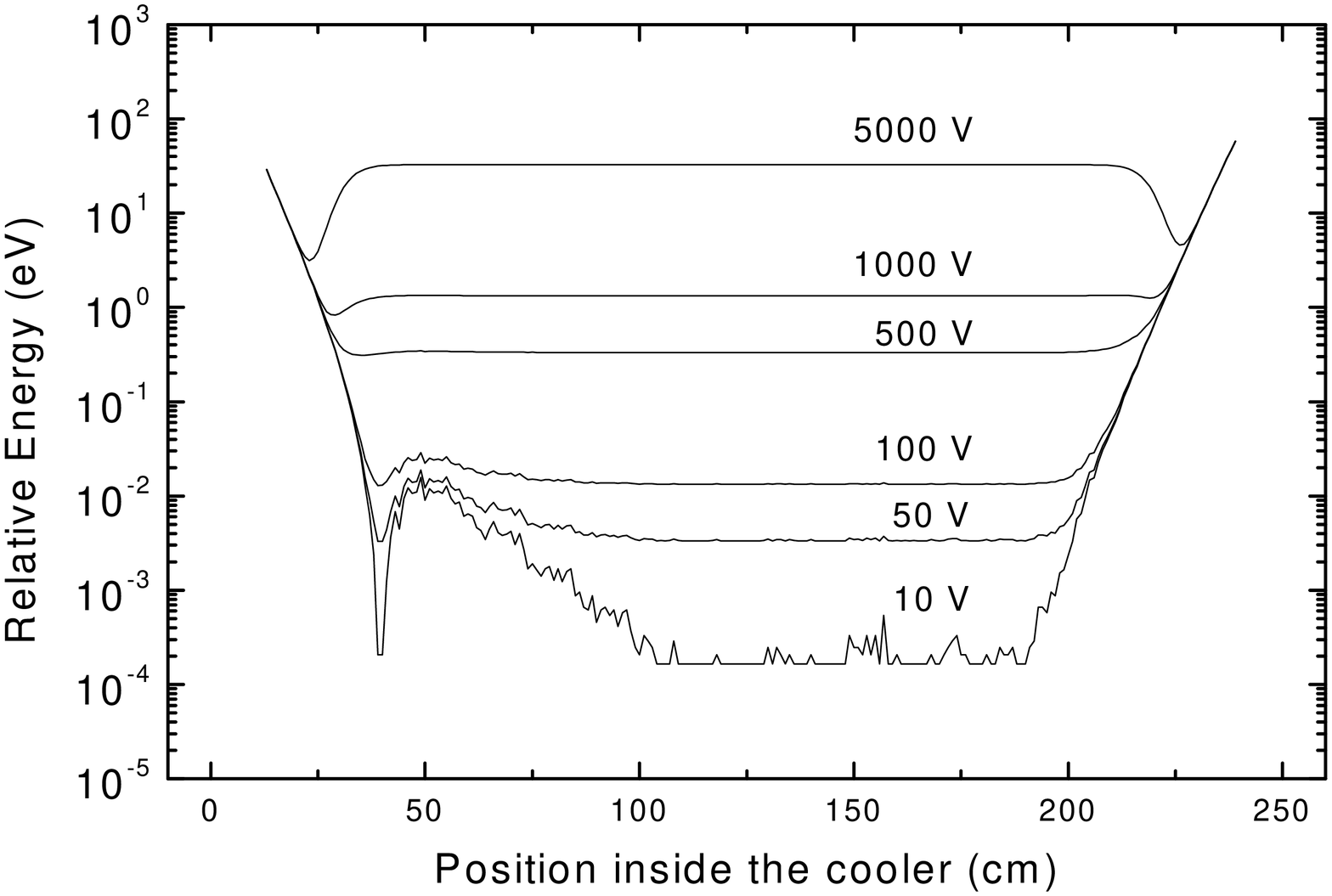}
\end{center}
\caption{Relative energies between electrons and ions along the
straight overlap
  section inside the cooler. The energies have been calculated according to
  Eq.~\ref{equ:rel_energy} taking into account the distributions of the electric
  potential and the angle $\phi$ (from Fig.~\ref{fig:potential}). The voltages of 10~V, 50~V,
  100~V, 500~V, 1000~V and 5000~V applied to the drift tubes correspond to $E_{rel}$ =
  1.7 $\times 10^{-4}$~eV, 0.003~eV, 0.013~eV, 0.33~eV, 1.3~eV and 32.7~eV. }\label{fig:energy}
\end{figure}

The right panel of Fig.~\ref{fig:potential} shows the distribution
of angles $\phi$ between electron trajectories and the ion beam
direction along the geometrical cooler axis. The electrons
strictly following the magnetic field lines, the non-zero angles
result from the measured transverse magnetic guiding field
components in this section. The measurement was restricted to a
length of 2.26~m (90~\% of the straight overlap section) fully
including the drift tube area. As one can see $\phi$ increases
rapidly at the edges of the measured range. Using steerer magnets
the electron beam can be shifted to a position minimizing the
influence of angular deviations. Both the distributions of the
electric potential and angle (Fig.~\ref{fig:potential}) can be
combined via Eq.~\ref{equ:rel_energy} into a distribution of
relative energies along the ion beam axis. Fig.~\ref{fig:energy}
shows the relative energies along the straight overlap section for
different voltages applied to the drift tubes. Obviously the
desired relative energies are only realized along a certain
fraction of the whole interaction length which, in addition,
depends on the energy. The measured rate coefficient at a given
relative energy $E_{meas}$ always contains contributions from
other relative energies according to:\\
\begin{equation}\label{equ:corr_rate}
  \alpha(E_{meas}) = \frac{1}{L} \int^L_0 dl~\alpha(E_{rel}(l))
\end{equation}
with $E_{rel}(l)$ being the relative energy at the position $l$
inside the cooler. According to Eq.~\ref{equ:corr_rate} the
correct rate coefficient can be obtained by a deconvolution which
is performed iteratively. In a first iteration step the measured
rate coefficient $\alpha(E_{meas}) = \alpha^{(0)}$ is inserted as
$\alpha(E_{rel}(l))$ in Eq.~\ref{equ:corr_rate}. Then the
difference $\Delta \alpha^{(0)}$ between the obtained result and
the measured rate is subtracted from $\alpha^{(0)}$. In a next
step the new $\alpha^{(1)} = \alpha^{(0)} - \Delta \alpha^{(0)}$
is likewise inserted in Eq.~\ref{equ:corr_rate} and the difference
$\Delta \alpha^{(1)}$ is calculated. The procedure is carried on
until in a step k the relative difference $\Delta
\alpha^{(k)}(E_{meas})/\alpha(E_{meas})$ is below $10^{-3}$ at all
measured energies. This is the case after only a few iteration
steps.

Although the relative statistical errors of the results presented
below amount to less than $1\%$ in the rate coefficient maximum,
the systematic uncertainty in the absolute recombination rate
coefficient has been estimated to be $\pm23\%$.

\section{Experimental results and discussion}
\label{sec:results}

\begin{figure}
\begin{center}
\epsfxsize=\columnwidth \epsfbox{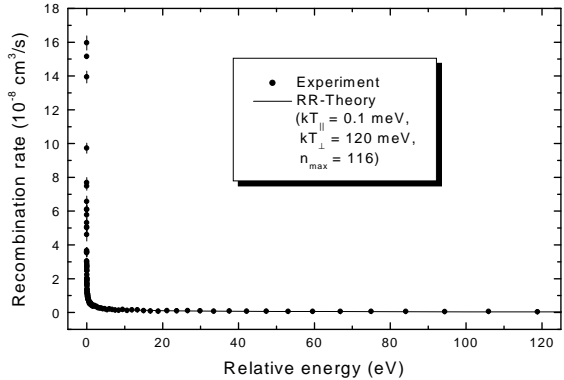}
\end{center}
  \caption{Measured absolute recombination rate of Bi$^{83+}$ (circles) with free electrons
  plotted against the relative energy between electrons and ions. The spectrum
  shows the typical shape of the RR peak with a continuous
  decrease of the rate
  with increasing energy. There is a very good agreement between the experimental data
  and RR theory from $E_{rel}$ = 15~meV to 125~eV. The theoretical curve
  follows from Eqs.~\ref{equ:sigmasum} and \ref{eq:faltung} using $kT_{||}$ = 0.1~meV,
  $kT_{\perp}$ = 120~meV and $n_{max}$ = 116 in the distribution function (Eq.~\ref{eq:fele}).}\label{fig:overview}

\end{figure}

In Fig.~\ref{fig:overview} the measured absolute rate coefficient
of Bi$^{83+}$ with free electrons is plotted versus the relative
energy from 0~eV to 125~eV. The spectrum shows the typical shape
of an RR rate coefficient with a maximum at $E_{rel}$ = 0~eV and a
continuous decrease for increasing relative energies.

For the comparison of the  measured rates with RR theory one has
to know the maximum principal quantum number $n_{max}$ and the
temperatures $T_{||}$ and $T_{\perp}$ (see sections
\ref{sec:theory} and \ref{sec:evaluation}). For the temperature
$T_{\perp}$ characterizing the transverse motion of the electrons
one has to assume $kT_{\perp}$ = 120~meV which corresponds to the
cathode temperature. For the longitudinal electron motion
$kT_{||}$ = 0.1~meV is inferred from the analysis of resonance
shapes in dielectronic recombination (DR) measurements with
lithium-like Bi$^{80+}$-ions performed during the same beamtime.
$n_{max}$ is determined by field ionization in the dipole magnet
which separates the parent beam and the recombined ions. A first
approximation of this value is the field ionization limit $n_F$,
which is obtained from
\begin{equation}
n_F= \left(7.3 \times 10^{10} {\rm Vm}^{-1} \frac{q^3}{F}\
\right)^{1/4}
\end{equation}
\cite{mueller2}, where q is the charge state of the ion and $F =
v_{i,||}B_{\perp}$ the motional electric field seen by the ions
with velocity $v_{i,||}$ in the transverse magnetic field
$B_\perp$ of the charge-analyzing magnet. In this context one also
has to account for the possibility that high Rydberg states can
decay to states below $n_F$, provided the ions have some time
between the recombination process and the arrival at the ionizing
electric field $F$. Therefore, a realistic estimate for the
cut-off is given by
\begin{equation}
n_{max} = \max(n_\gamma,n_F),
\end{equation}
where $n_\gamma$ denotes the maximum principal quantum number of
Rydberg states which decay before the recombined ions arrive at
the analyzing magnet and thus are saved from field ionization. A
crude estimate of $n_\gamma$ based on a number of assumptions on
the population and the decay of excited states is obtained from a
numerical solution of the following equation \cite{wolf2}\\
\begin{eqnarray}
\label{equ:n_cut}
 n_\gamma  & = & Z^{4/5} \{2.142 \times 10^{10}
s^{-1} t_F\left[\kappa(n_\gamma)\right]^2 \nonumber \\ & &
\left[-0.04+\ln(n_\gamma)-\ln\kappa(n_\gamma)\right]\}^{1/5},
\nonumber \\ \kappa(n_\gamma)  & = & 1.6 + 0.018n_\gamma.
\end{eqnarray}
The time of flight $t_F$ of the ions between the recombination act
and the arrival at the analyzing magnet influences the survival
probability of ions in high Rydberg states. In the derivation of
Eq.~\ref{equ:n_cut} cascade processes were neglected. They should
not be of large influence considering the decay rates in
undisturbed hydrogenic systems. A more problematic assumption
underlying Eq.~\ref{equ:n_cut} is that the fields seen by the ions
during their flight time are considered not to change the {\it nl}
distribution of states as it results from the initial population
by radiative recombination. Changes of decay rates by Stark mixing
in the fields are not accounted for but probably deserve further
attention. Also, instead of a sharp cut-off at $n_\gamma$ or $n_F$
a distribution of field ionization probabilities \cite{mueller2}
around $n_\gamma$ or $n_F$, respectively, would be more realistic,
however, these cannot easily be calculated without further
assumptions. In addition, the influences of the experimental
conditions providing electric fields via $\vec{v} \times \vec{B}$
contributions inside the interaction area, the toroidal fields and
the dipole correction magnets after the cooler have to be
considered.

\begin{figure}
\begin{center}
\epsfxsize=\columnwidth \epsfbox{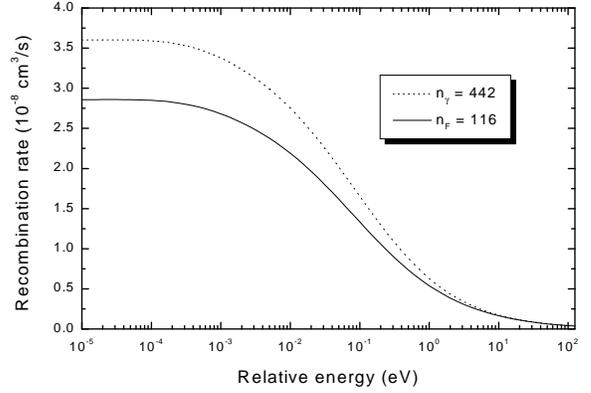}
\end{center}
  \caption{Calculated rate coefficients for $n_{max} = n_{\gamma}$ = 442 and $n_{max} = n_F$ = 116.
  For both curves electron beam temperatures $kT_{||}$ = 0.1~meV and $kT_{\perp}$ = 120~meV have been
  used.}\label{fig:nmax}
\end{figure}

In the present experiment $n_F$ and $n_\gamma$ have been
calculated to be 116 and 442, respectively. A comparison of both
RR rate curves resulting from Eqs.~\ref{equ:sigmasum} and
\ref{eq:faltung} is shown in Fig.~\ref{fig:nmax}. As expected the
theoretical rate for $n_{max}$ = $n_\gamma$ = 442 (dotted line)
lies above the curve with $n_{max}$ = $n_F$ = 116 (solid line).
Comparing the experimental and the theoretical rate coefficients
the curve calculated for $n_F$ = 116 shows a very good agreement
with the experimental data for relative energies from $E_{rel}$ =
15~meV to 125~eV (Fig.~\ref{fig:overview}). However, one can also
obtain a good agreement for $n_{\gamma}$ = 442 if one assumes a
higher transverse electron temperature $kT_{\perp}$ = 250~meV.
Such an interdependency between $n_{max}$ and $kT_{\perp}$ has
also been found in experiments with bare Cl$^{17+}$ and
C$^{6+}$-ions at the TSR in Heidelberg \cite{hoffknecht2,gwinner}.
Because of the impossibility to accurately obtain both parameters
from a fit of the theoretical RR curve to the experimental
spectrum and in view of the rather crude estimation of $n_{max}$
we deliberately choose $n_{max}$ = $n_F$ = 116 throughout the rest
of this paper. This implies a reasonable choice for the transverse
temperature of $kT_{\perp}$ = 120~meV, which is in accordance with
the cathode temperature. That same temperature is also suggested
by the accompanying DR measurements with Bi$^{80+}$-ions. In this
discussion the longitudinal temperature $T_{\parallel}$ does not
play a role since it has very little influence on the RR rate
coefficient.

\begin{figure}
\begin{center}
\epsfxsize=\columnwidth \epsfbox{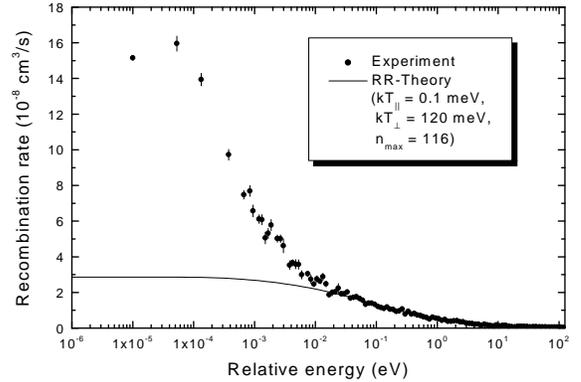}
\end{center}
 \caption{Comparison between the measured rate coefficient for Bi$^{83+}$ and RR theory. At $E_{rel}$
  = 0~eV a theoretical rate coefficient of 2.8$\times10^{-8}$~cm$^3$~s$^{-1}$ has been obtained using
  $kT_{||}$ = 0.1~meV, $kT_{\perp}$ = 120~meV and $n_{max}$ = 116. The experimental rate of
  1.5$\times10^{-7}$~cm$^3$~s$^{-1}$ exceeds this value by a factor of 5.2. }\label{fig:enhancement}
\end{figure}

In Fig.~\ref{fig:enhancement} the experimental and theoretical
data of Fig.~\ref{fig:overview} are shown again using
 a logarithmic energy scale in order to focus on a comparison at
very low energies.
 The shape of the experimental spectrum with an
additional increase towards low energies $\leq$ 15~meV is typical
for low energy recombination measurements. At E$_{rel}$ = 0~eV we
obtained a maximum rate coefficient of 1.5$\times
10^{-7}$~cm$^{3}$~s$^{-1}$ exceeding the theoretical rate of
2.9$\times 10^{-8}$~cm$^3$~s$^{-1}$ by a factor of 5.2. As already
mentioned in the introduction this rate enhancement phenomenon has
also been observed at other facilities, however, the
Bi$^{83+}$-experiment provides the first quantitative
determination of the rate enhancement factor $\epsilon$ for a bare
ion with $Z \geq 18$. As mentioned above, for the light ions
He$^{2+}$, N$^{7+}$, Ne$^{10+}$ and Si$^{14+}$ a $Z^{2.8}$
dependence of $\Delta \alpha$ = $\alpha_{exp} - \alpha_{theo}$ was
found in Stockholm \cite{gao1}. This scaling cannot be directly
confirmed for Bi$^{83+}$. In order to obtain an agreement with the
$Z^{2.8}$-scaling the measured rate coefficient for Bi$^{83+}$
would have to be more than 400~\% higher which is beyond the
experimental uncertainty. However, one has to be careful with
comparing results from different facilities since the experimental
conditions vary drastically. Apart from  the extremely high
electron energy of 162~keV and the extremely high nuclear charge
in the present case the influence of the experimental parameters
$kT_{\perp}$, $kT_{\parallel}$ and the magnetic guiding field B
has to be considered. It is known from previous experiments with
F$^{6+}$ and C$^{6+}$ ions \cite{gwinner} that the excess rate
$\Delta \alpha$ scales as $(kT_{||})^{-1/2}$ and as
$(kT_{\perp})^{-1/2}$. Using the present temperatures in
comparison with the Stockholm conditions the excess rate found in
the present experiment has to be multiplied by a factor of
approximately $\sqrt{120~mev/10~meV} \cdot \sqrt{0.1~meV/0.12~meV}
\approx$ 3.2 in order to normalize it to the Stockholm conditions.
This removes part of the discrepancy mentioned above. The magnetic
field dependence of $\Delta \alpha$ found in similar experiments
with other ions would, however, reduce the normalization factor
again. To little is known to date about the exact numbers for such
a normalization.

\begin{figure}
\begin{center}
\epsfxsize=\columnwidth \epsfbox{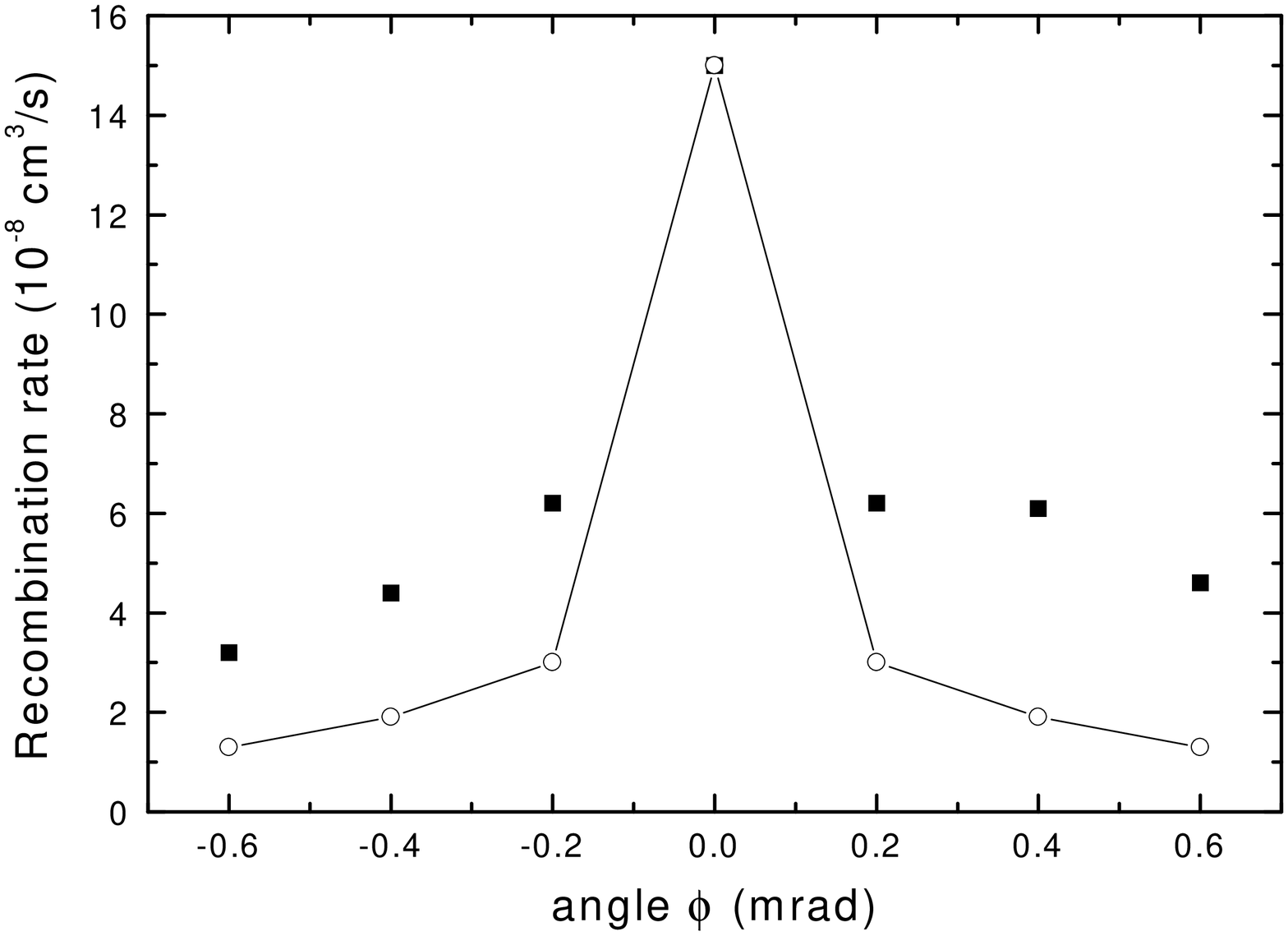}
\end{center}
  \caption{Maximum recombination rates at $E_{rel}$ = 0~eV for different angles $\phi$ set between electron and ion beam.
  The open circles denote the values expected on the basis of the ($\phi$ = 0~mrad)-spectrum. The measured rate coefficients
  are represented by the squares. The differences at angles larger than 0~mrad a partially due to the small
  misalignments in the magnetic field (Fig.~\ref{fig:potential}) as well as to the betatron oscillations.}\label{fig:angles}
\end{figure}

As mentioned already in section \ref{sec:experiment} the alignment
of the beams has been carefully optimized before starting the
recombination experiment. During the measurement  we artificially
introduced an angle $\phi$ between the beams in order to check the
obtained settings. This was implemented by superimposing in the
interaction region a defined transverse (horizontal) magnetic
field $B_x$ in addition to the unchanged longitudinal field $B_z$
along the ion beam direction. Fig.~\ref{fig:angles} shows the
maximum recombination rate at $E_{rel}$ = 0~eV for different
angles $\phi$ from -0.6~mrad to 0.6~mrad in the horizontal plane.
At $\phi$ = 0~mrad the maximum recombination rate is obtained. The
open circles in Fig.~\ref{fig:angles} denote the expected rates
for the selected angles. They have been determined by taking
recombination rates from the ($\phi$ = 0~mrad)-spectrum at the
minimum relative energies possible at the corresponding angle (see
eq.~\ref{equ:rel_energy}). The squares in Fig.~\ref{fig:angles}
represent the measured rate coefficients at the minimum relative
energies $E_{rel}$ accessible at the selected angles. All of them
are lying above the expected value. Such a behaviour is already
known from other experiments at the ESR. Obviously the ion beam
reacts on the introduction of the transverse magnetic field
component, Lorentz and cooling forces appear to minimize the
effect of the change. In any case, the distribution shows a rather
symmetric progression with the maximum at $\phi$ = 0~mrad
confirming the accurate adjustment of the cooler.

\begin{figure}
\begin{center}
\epsfxsize=\columnwidth \epsfbox{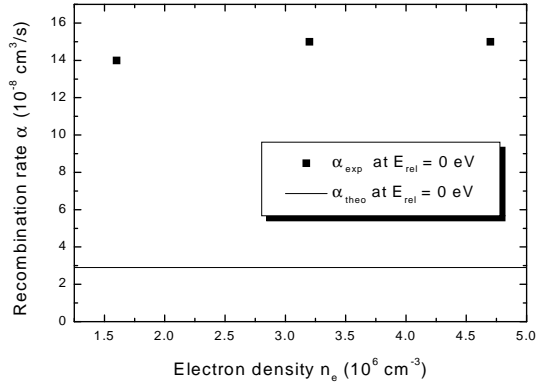}
\end{center}
  \caption{Influence of the electron density on the recombination of Bi$^{83+}$. The squares represent
  the measured rate coefficient at $E_{rel}$ = 0~eV plotted against the electron density. The solid line
  shows the theoretical RR rate coefficient at $E_{rel}$ = 0~eV calculated with $kT_{||}$ = 0.1~meV and
  $kT_{\perp}$ = 120~meV.}\label{fig:density}
\end{figure}

\begin{figure}
\begin{center}
\epsfxsize=\columnwidth \epsfbox{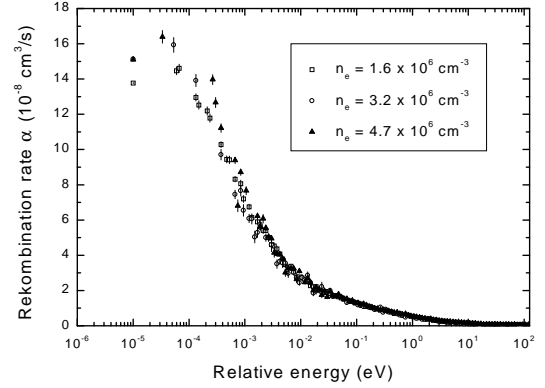}
\end{center}
  \caption{Measured absolute rate coefficients for Bi$^{83+}$ for different electron
  densities. The shapes of the spectra are practically identical. Greater
  fluctuations of the recombination rate for $n_e =
  4.7\times$10$^{6}$~cm$^{-3}$ are due to the measurement procedure.
  For this electron density a random distribution of
  the drift tube voltages has been used instead of a continuous
  increase of the voltage for each measurement step. Comparisons
  of spectra measured with both procedures show the same
  progressions but the spectra obtained with the random
  distribution are not so smooth.
  }\label{fig:shape}
\end{figure}

In order to investigate the influence of the electron density on
the recombination rate we performed recombination measurements for
three different densities $n_e$ = 1.6$\times$10$^{6}$~cm$^{-3}$,
$n_e = 3.2\times10^{6}$~cm$^{-3}$ and
4.7$\times$10$^{6}$~cm$^{-3}$. In Fig.~\ref{fig:density} the rate
coefficient at E$_{rel}$= 0~eV is plotted against the electron
density n$_{e}$. The solid line represents the theoretical rate
coefficient calculated with kT$_{||}$= 0.1~meV and kT$_{\perp}$=
120~meV at E$_{rel}$= 0~eV. There is a small difference between
the maximum rate coefficient $\alpha_{max} =
1.4\times10^{-7}$~cm$^{3}$~s$^{-1}$ for $n_e$ =
1.6$\times$10$^{6}$~cm$^{-3}$ and $\alpha_{max} =
1.5\times10^{-7}$~cm$^{3}$~s$^{-1}$ for the two higher densities
but this deviation is within the experimental uncertainty. In
addition Fig.~\ref{fig:shape} shows that the shape of the spectra
is equal for all densities indicating identical temperatures in
the measurements. Therefore it can be concluded that there is no
significant influence of the electron density on the recombination
rate. This observation is in accordance with findings at the
CRYRING \cite{gao2}, the TSR \cite{hoffknecht2} and the GSI single
pass electron target \cite{hoffknecht1}. The lack of any density
dependence rules out TBR (eq.~\ref{equ:tbr}) as a possible
mechanism leading to enhanced recombination rates at low energies.
With a significant contribution of TBR to the observed rates one
would expect an increase of the recombination rate with increasing
electron density in contrast to all experimental observations. In
the context of storage rings TBR has been discussed in some detail
by Pajek und Schuch \cite{pajek}. They found theoretically that
TBR effectively populates high Rydberg states of the ion where the
electrons are very weakly bound. As mentioned above such ions are
reionized in the dipole magnet and therefore do not contribute to
the measured recombination rate.

\begin{figure}
\begin{center}
\epsfxsize=\columnwidth \epsfbox{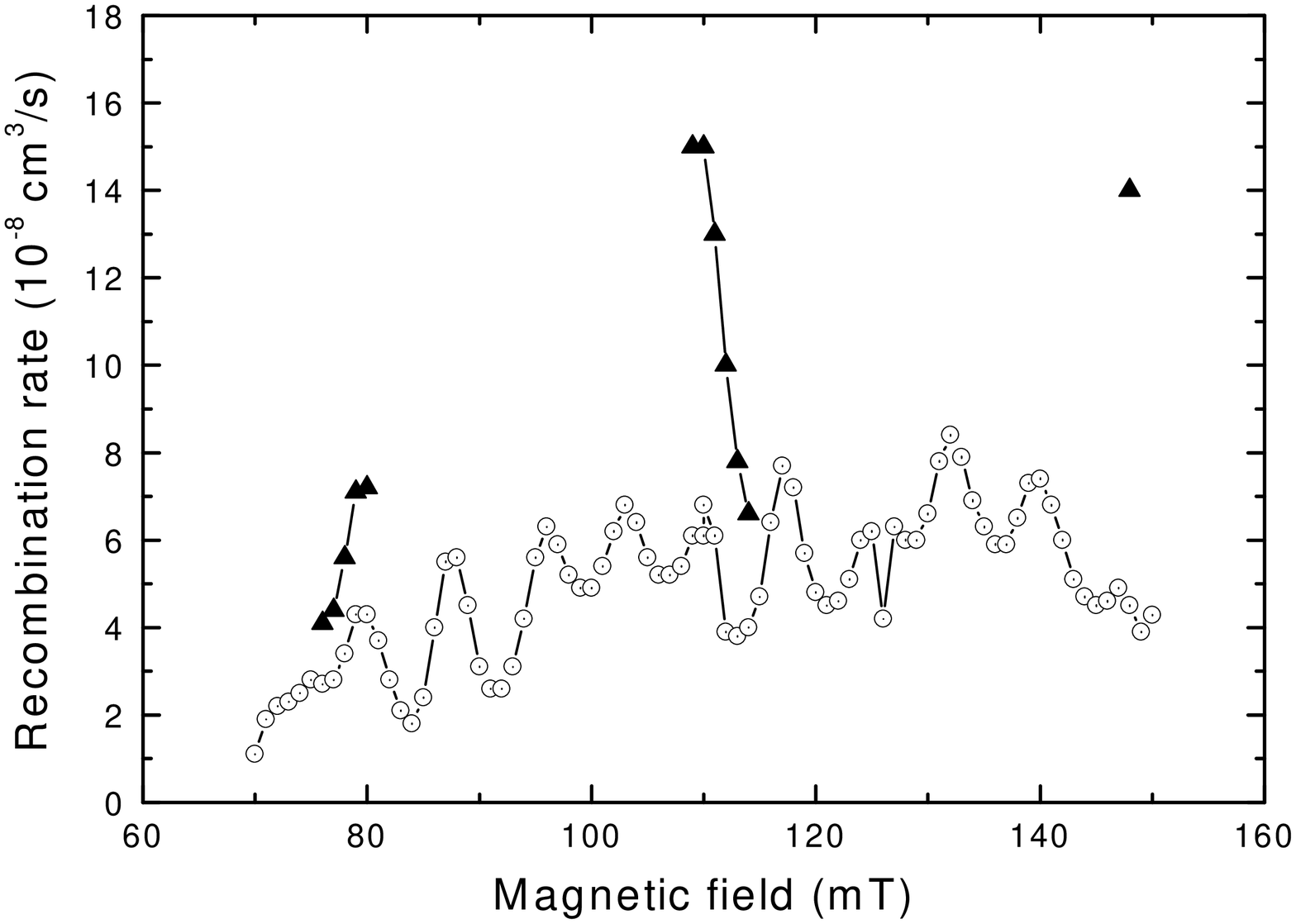}
\end{center}
  \caption{Measured maximum recombination rate at $E_{rel}$ = 0~eV versus the magnetic field. The open
  circles represent the uncorrected recombination rate (see also Fig.~\ref{fig:rates}). Complete recombination
  measurements with all corrections were performed for 12 selected field strengths
  indicated by the full triangles. }\label{fig:magnet1}
\end{figure}

In a next stage of our experiment we also varied the magnetic
guiding field of the electron beam between 70~mT and 150~mT in
steps of 1~mT. The standard field strength used for the previous
measurements was 110~mT. In contrast to the more careful
adjustments of the magnetic field at the TSR
\cite{hoffknecht3,gwinner} which were accompanied by measurements
of the cooling force and the beam profiles in order to preserve
the beam quality no other cooler setting beside the magnetic field
was changed at the ESR. This procedure was motivated by an earlier
experiment of the ESR cooler group \cite{steck} showing an
oscillation of the recombination rate for 310~MeV/u U$^{92+}$-ions
induced by small changes of the magnetic field.  The new results
obtained for 295.3~MeV/u Bi$^{83+}$ are shown in
Fig.~\ref{fig:magnet1} where the maximum recombination rate at
$E_{rel}$ = 0~eV is plotted versus the magnetic field strength
$B_{||}$. Since the measurement of a complete recombination
spectrum is very time-consuming only the recombination rate at
cooling has been recorded for each magnetic field setting. For
these data points a background subtraction and corrections due to
the potential and angle distributions inside the cooler were not
possible. Therefore these recombination rates represented by the
open circles in Fig.~\ref{fig:magnet1} can only display the
qualitative dependance on the magnetic field.

\begin{figure}
\begin{center}
\epsfxsize=\columnwidth \epsfbox{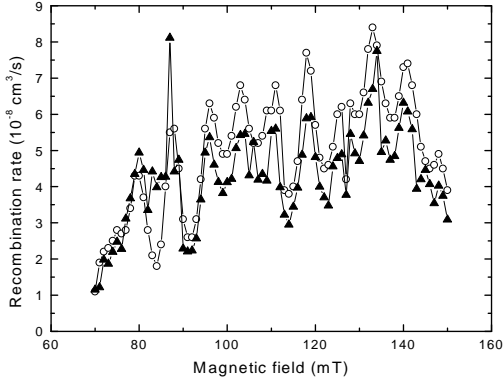}
\end{center}
  \caption{Comparison of recombination rates at $E_{rel}$ = 0~eV obtained via two different methods.
  The open circles show the recombination rate calculated from the counting rate of recombined particles. These data
  do not include the background subtraction and corrections due to the potential and angle distribution inside the
  cooler. Therefore it can only give a qualitative overview.
  The full triangles represent rate coefficients obtained via the storage lifetime of the Bi$^{83+}$ beam in the ring.
  There is a good overall agreement between the different approaches.}\label{fig:rates}
\end{figure}

A complementary method of determining the recombination rate at
cooling can be applied by analyzing the lifetime of the Bi$^{83+}$
beam in the ring. Fitting an exponential curve $I(t) = I_0 \cdot
\exp(-t/ \tau)+I_{Off}$ to the decay of the ion current stored in
the ring with $I_0$ being the ion current at the beginning and
with $I_{Off}$ representing the offset of the ion current
transformer, one can extract the storage lifetime $\tau$ of the
Bi$^{83+}$-beam in the ring. Assuming that electron-ion
recombination in the cooler is the only loss mechanism for stored
ions one can calculate the corresponding recombination rate
$\alpha_l = 1/(\tau n_{eff})$. $n_{eff}$ is the effective electron
density which is $n_e$ times the ratio of the interaction length
and the ring circumference. Fig.~\ref{fig:rates} shows a
comparison of the rates obtained with the different methods
resulting in a good agreement of the data.

 A Fourier analysis of
the experimental data reveals an "oscillation period" of the
recombination rate of 7.6~mT. Recent measurements of the ESR
cooler group with 300 MeV/u Kr$^{36+,35+}$-ions show that this
value can be influenced by a variation of the magnetic fields of
the toroids (see Fig.~\ref{fig:cooler}) moving the electron beam
into and out of the interaction area . Due to the Lorentz force
the electrons are moving on helical trajectories through the
cooler. It has been observed by the ESR group that the
"oscillation period" corresponds to a change of the magnetic field
that allows the electrons one more turn inside the toroid. In
order to find an explanation for these observations one has to
take a precise look at the measured recombination spectra.
Therefore we performed complete recombination measurements with
all corrections for 12 selected field strengths. The results are
represented by the full triangles in Fig.~\ref{fig:magnet1}. The
maximum recombination rates  at $E_{rel}$ = 0~eV obtained in these
measurements show the same progression as the open circles.

\begin{figure}
\begin{center}
\epsfxsize=\columnwidth \epsfbox{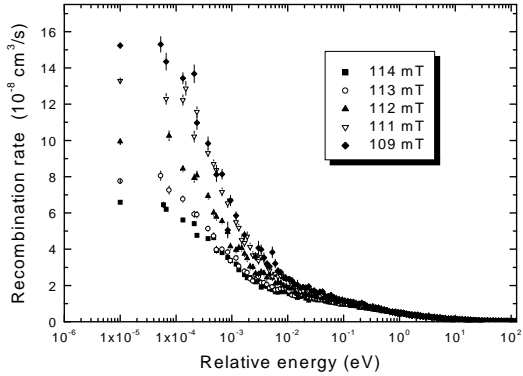}
\end{center}
 \caption{Comparison of Bi$^{83+}$ recombination spectra for magnetic fields between B = 109~mT
  and 114~mT. For relative energies $E_{rel} \geq$ 1~eV the shapes of the spectra are practically
  identical. At energies below 1~eV there are significant differences depending on the
  magnetic field.}\label{fig:magnet2}
\end{figure}

\begin{figure}
\begin{center}
\epsfxsize=\columnwidth \epsfbox{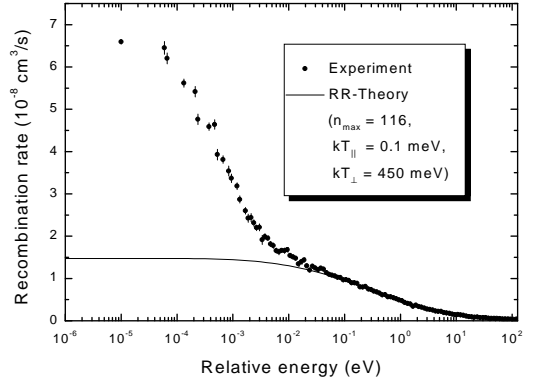}
\end{center}
  \caption{Comparison between the measured absolute rate coefficient for Bi$^{83+}$
  at the oscillation minimum at B = 114~mT and RR theory. In order to describe the experimental progression
  the theoretical curve was calculated with a transverse temperature $kT_{\perp}$ = 450~meV. The
  longitudinal temperature $kT_{||}$ = 0.1~meV and the maximum principal quantum number $n_{max}$ = 116
  were the same as before. }\label{fig:magnet3}
\end{figure}

In Fig.~\ref{fig:magnet2} recombination spectra for different
magnetic fields between B = 109~mT and 114~mT are shown in the
energy range from $E_{rel}$ = 0~eV to 125~eV. For $E_{rel} \geq$
1~eV the measured rate coefficients are practically identical. At
energies below 1~eV the recombination rates show significant
differences depending on the magnetic field strength. However, the
rate enhancement is always there which one can see in
Fig.~\ref{fig:magnet3} where the recombination rate for the
oscillation minimum at B = 114~mT is compared with RR theory. In
order to describe the experimental data a transverse electron
temperature of $kT_{\perp}$ = 450~meV was applied which is nearly
a factor of 3 higher than the one obtained by a fit of RR theory
to the experimental data at the standard magnetic field strength
of B = 110~mT. From the measured rate at $E_{rel}$ = 0~eV of
$\alpha_{exp}$ = 6.6$\times10^{-8}$~cm$^3$~s$^{-1}$ and the
theoretical value at $E_{rel}$ = 0~eV of $\alpha_{theo}$ =
1.5$\times10^{-8}$~cm$^3$~s$^{-1}$ one calculates a rate
enhancement factor $\epsilon = \alpha_{exp}/\alpha_{theo}$ = 4.4.

\begin{figure}
\begin{center}
\epsfxsize=\columnwidth \epsfbox{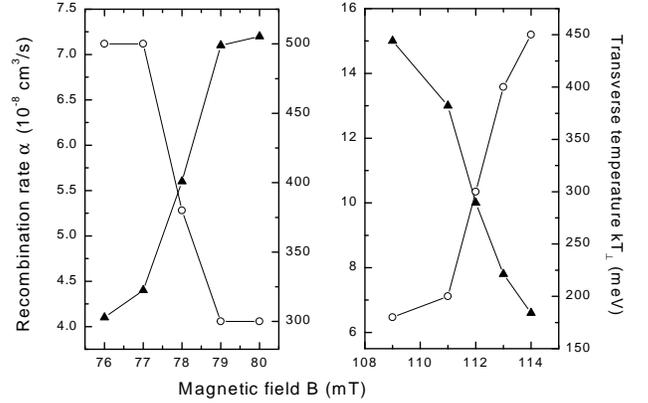}
\end{center}
  \caption{Comparison of the measured maximum rate coefficients (full triangles) for different magnetic fields
  and the transverse electron temperatures (open circles) inferred from fits to the corresponding recombination
  spectra at energies beyond 0.1~eV. The left panel represents the data for magnetic fields between B = 76~mT and 80~mT.
  As already displayed in Fig.~\ref{fig:magnet1} the rate coefficient increases with increasing
  magnetic field. The transverse electron temperature shows the opposite behaviour and decreases
  with increasing magnetic field. Thus the magnitude of the maximum rate coefficient at $E_{rel}$ =
  0~eV is apparently strongly related to the transverse electron temperature. This relationship is also
  present in the right panel in which the data for magnetic fields between B = 109~mT and
  114~mT is presented.}\label{fig:magnet4}
\end{figure}

Analysing all the complete recombination spectra measured for
different magnetic fields a connection between the maximum rate
coefficient at $E_{rel}$ = 0~eV and the adapted transverse
electron temperature $kT_{\perp}$ becomes obvious which is
documented in Fig.~\ref{fig:magnet4}. In the left panel
representing the data for magnetic fields between 76~mT and 80~mT
the measured maximum rate coefficient at $E_{rel}$ = 0~eV (full
triangles) increases with increasing magnetic field whereas the
transverse electron temperature (open circles) adapted to the
corresponding recombination spectra decreases in this region. In
the right spectrum monitoring the same data for magnetic fields
between 109~mT and 114~mT one can see a decrease of the maximum
rate coefficient at $E_{rel}$ = 0~eV accompanied by an increase of
the transverse electron temperature $kT_{\perp}$. Therefore there
seems to be a relationship between the rate coefficient and the
transverse electron temperature (or: the electron beam quality)
which would easily explain the periodic reductions of the
recombination rate at $E_{rel}$ = 0~eV. Nevertheless it should be
pointed out that this interpretation does not explain the observed
general enhancement of the rate coefficient at $E_{rel}$ = 0~eV. A
further theoretical analysis should especially focus on the
possible influence of the electron energy since in experiments
with lithium-like 97.2~MeV/u Bi$^{80+}$ ions (corresponding to
53.31~kV cooling voltage instead of 162~kV for 295.3~MeV/u) during
the same beam-time the oscillations did not appear. This is in
agreement with the results obtained at the ESR and other storage
rings in experiments at low ion energies. There, oscillations of
the recombination rate at cooling have not been observed.

\section{Conclusions}
\label{sec:conclusions} The recombination of bare Bi$^{83+}$ ions
with  free electrons has been studied at the GSI Experimental
Storage Ring (ESR) in Darmstadt. Within the experimental
uncertainty we found very good agreement between the measured rate
coefficient and the theory for radiative recombination (RR) for
energies from E$_{rel}$ = 15~meV to 125~eV. At very low
center-of-mass energies between ions and electrons, however, the
measured rate exceeds the theoretical predictions by a factor of
5.2. This first rate enhancement result for a bare ion with $Z$
very much greater than 18 does not follow a $Z^{2.8}$ dependence
of the enhancement found in an experiment with the light ions
He$^{2+}$, N$^{7+}$, Ne$^{10+}$ and Si$^{14+}$. Therefore
additional measurements in the mid-Z range are necessary in order
to obtain more information about the influence of $Z$ on the
recombination enhancement phenomenon. The increase of the electron
density in the interaction area from $n_e$ =
1.6$\times$10$^{6}$~cm$^{-3}$ to 4.7$\times$10$^{6}$~cm$^{-3}$
appears to have no significant effect on the recombination rate. A
variation of the magnetic field from 70~mT to 150~mT revealed a
strong dependence of the recombination rate at low energies on
this parameter. The observed oscillations of the maximum
recombination rate at $E_{rel}$ = 0~eV confirmed previous
observations with bare U$^{92+}$ ions. Comparing the recombination
spectra with RR theory one finds strong variations of the
transverse electron temperature connected to the oscillations of
the recombination rate. In future experimental and theoretical
studies this relationship has to be investigated in more detail.
Finally, we want to emphasize, that the observed recombination
rate enhancement significantly reduces the lifetime of ion beams
in storage rings during the electron cooling procedure. At the
present ion energies, recombination in the cooler by far dominates
over all factors influencing the beam lifetime. The recombination
rate enhancement hence reduces the beam lifetime by approximately
a factor 5 as compared to the assumption of pure RR.
\section{Acknowlegments}
\label{sec:acknowledments} The Giessen group gratefully
acknowledges support for this work through contract GI M\"{U}L S
with the Gesellschaft f\"{u}r Schwerionenforschung (GSI),
Darmstadt, and by a research grant (number 06 GI 848) from the
Bundesministerium f\"ur Bildung und Forschung (BMBF), Bonn.

\end{document}